\documentclass{article}
\usepackage{spconf,amsmath,graphicx}
\usepackage{booktabs}
\usepackage{multirow}
\usepackage{array}
\usepackage{makecell}
\usepackage{hyperref}
\usepackage[
backend=biber,
style=ieee,
maxbibnames=3,
maxcitenames=3,
doi=false,isbn=false,url=false,eprint=false
]{biblatex}
\hypersetup{
     colorlinks=true,
     citecolor=blue,
}
\defbibheading{bibliography}[\refname]{}
\renewcommand*{\bibfont} 

\title{Electrolaryngeal Speech Intelligibility Enhancement Through Robust Linguistic Encoders}
\name{\vspace{-0.4em}\begin{tabular}{c}Lester Phillip Violeta$^{1}$, Wen-Chin Huang$^{1}$, Ding Ma$^{1}$, Ryuichi Yamamoto$^{1}$, \\ Kazuhiro Kobayashi$^{1,2}$, Tomoki Toda$^{1}$ \end{tabular}}
\address{$^1$Nagoya University, Japan, $^{2}$TARVO, Inc., Japan
}

\begin{document}
\ninept
\maketitle
\vspace{7.4em}
\begin{abstract}
We propose a novel framework for electrolaryngeal speech intelligibility enhancement through the use of robust linguistic encoders. Pretraining and fine-tuning approaches have proven to work well in this task, but in most cases, various mismatches, such as the speech type mismatch (electrolaryngeal vs. typical) or a speaker mismatch between the datasets used in each stage, can deteriorate the conversion performance of this framework. To resolve this issue, we propose a linguistic encoder robust enough to project both EL and typical speech in the same latent space, while still being able to extract accurate linguistic information, creating a unified representation to reduce the speech type mismatch. Furthermore, we introduce HuBERT output features to the proposed framework for reducing the speaker mismatch, making it possible to effectively use a large-scale parallel dataset during pretraining. We show that compared to the conventional framework using mel-spectrogram input and output features, using the proposed framework enables the model to synthesize more intelligible and naturally sounding speech, as shown by a significant 16\% improvement in character error rate and 0.83 improvement in naturalness score.
\end{abstract}
\section{Introduction}
\label{sec:intro}
Voice conversion (VC) \cite{vc-intro}, the task known as changing the speaker information while keeping linguistic information unchanged, has had rapid improvements in the age of deep learning. One of its sub-applications, intelligibility enhancement \cite{project-euphonia, parrotron, extending-parrotron}, has made way for atypical speakers to regain the ability to speak like typical speakers. Atypical speakers have difficulties in producing phoneme sounds and speak at a slower rate, making daily communication a tedious task for them. One type of atypical speech, electrolaryngeal (EL) speech, is produced by speakers diagnosed with a disrupted larynx, the organ responsible for generating the source excitation. While an electrolarynx \cite{electrolarynx} is used as a replacement for the larynx, the resulting speech becomes unnatural due to the electrolarynx producing robotic-like source excitation and being unable to produce natural pitch variation. For pitch-based languages like Japanese, changing the pitch throughout a sentence along with the use of voiced/unvoiced sounds, is essential to infer the meaning of different words, making this an important task.

Several previous works in intelligibility enhancement have found that an effective solution is to first learn the alignments between typical and atypical speech through a parallel dataset. Since EL speakers speak at a slower rate and are unable to pronounce some phoneme sounds, learning the alignment between the two is essential in this task. For example, \cite{vtn-el-first} does this by using a strong sequence model such as a Transformer \cite{transformer, vtn_journal}. Due to data scarcity and the data-hungry nature of Transformer-based models, several works \cite{extending-parrotron, conformer-parrotron} have emphasized the effectiveness of pretraining on a large-scale typical speech dataset and fine-tuning it on a small-scale atypical speech dataset. However, a major problem in this naive pretraining and fine-tuning framework is that the typical and EL speech types are vastly different from each other. Thus, although a simple pretraining and fine-tuning approach brings in improvements, there is a performance ceiling in such an approach. Our previous work \cite{el_ma} resolved this by observing that fine-tuning first with large-scale synthetic speech can effectively soften the mismatch between the speech types and speakers, making the pretraining and fine-tuning approach more effective. However, there is still a lot of room for improvement in further reducing the speech type and speaker mismatches, as the synthesis performance is still far from human-level speech.

We resolve the speech type and speaker mismatch issues encountered in pretraining and fine-tuning approaches by introducing a new framework which uses recognition, alignment, and synthesis modules. Specifically, we use strong recognition modules containing dense linguistic information (such as bottleneck \cite{liu2021bnfvc} and HuBERT \cite{hubert, hubert-soft} features) as input and output features of the alignment module, effectively allowing the alignment module to focus on solely learning linguistic features. With the recognition module extracting pure linguistic-related information, we effectively remove the speech type and speaker mismatches occurring between each stage in pretraining and fine-tuning, resulting in better performance compared to the baseline. Moreover, with the use of a Diffusion-based \cite{ho2020denoising} synthesis decoder to generate the target speaker mel-spectrogram from the HuBERT output features, we shift the burden of synthesizing to a target speaker to this module due to its strong generation capabilities, improving the generation quality of the waveform. Finally, this proposed framework optimizes the use of parallel VC pretraining to further improve performance. Our contributions are summarized as follows:
\begin{itemize}
  \item We propose a novel framework for electrolaryngeal speech intelligibility enhancement, composed of recognition, alignment, and synthesis modules. We show that using this framework can synthesize speech with a 16\% CER improvement and a 0.83 higher naturalness score compared to the baseline. 
  \item We resolve the speech type mismatch issues by developing a linguistic encoder robust to both EL and typical speech types. Through a unified representation being used as inputs, the alignment module can focus on solely modeling the linguistic features, resulting in significantly more intelligible speech.
  \item We show the other important components of the framework, such as using HuBERT output features and using parallel VC pretraining in ablation studies.
\end{itemize}

\section{Conventional framework}
\label{baseline}
We use our previous work \cite{el_ma} as our baseline, which uses the Transformer \cite{transformer, vtn_journal} to transform the mel-spectrogram of an EL speech utterance into a mel-spectrogram of a typical speech utterance. 
A pretraining technique using text-to-speech (TTS) and autoencoder (AE) was used to efficiently learn linguistic information from a large-scale typical speech data. The process is done by first training a TTS model with a large-scale dataset. Then, an AE-style pretraining is conducted by using the decoder of the TTS model as initialization parameters, and reconstructing the target speaker by also using it as inputs. Here, the decoder parameters are frozen, such that the encoder is efficiently pretrained. The network is first fine-tuned on the parallel synthetic EL and typical speech as we found that fine-tuning first on synthetic EL speech (even with lots of mispronunciations in synthesis) softens the speech type and speaker mismatches when fine-tuning. Then, network is fine-tuned on the target EL and typical speech data. Moreover, since the TTS pretraining technique uses text information as inputs and models strong linguistic information \cite{vtn_journal}, such a speaker-independent pretraining style was beneficial in reducing the speech type and speaker mismatches when fine-tuning.

Although bringing in large improvements, the framework is still limited as it is still far from human-level speech. The main problem in this framework is that the mel-spectrograms contain a lot of information related to the speech type and speaker, degrading the performance due to the speech type and speaker mismatches between the datasets used in fine-tuning in each stage. One way to resolve this issue is by using linguistic encoders, which has shown success in several works in speech synthesis \cite{vc-intro, liu2021bnfvc}. By using a linguistic encoder to extract dense linguistic information from speech and using these as the input and output features, the focus during conversion can be shifted to the linguistic-related features, reducing the speech type and speaker mismatches. This has been applied to intelligibility enhancement where works such as \cite{robust_encoder_atypical} use a fine-tuned automatic speech recognition (ASR) model on the atypical speech; however, this approach cannot be directly applied to the pretraining and fine-tuning framework used in the majority of works. Although the ASR model can effectively extract linguistic features from the atypical speech, the pretraining on the large-scale typical speech dataset becomes less effective as the ASR model fine-tuned on the atypical speech cannot properly decode typical speech. In this work, we further investigate how to develop such a robust linguistic encoder and its observe its performance when used as input and output features for the Transformer network.

\vspace{-1.0em}
\section{Proposed framework}
\label{sec:format}
\begin{figure}[t!]
  \centering
  \includegraphics[width=7cm]{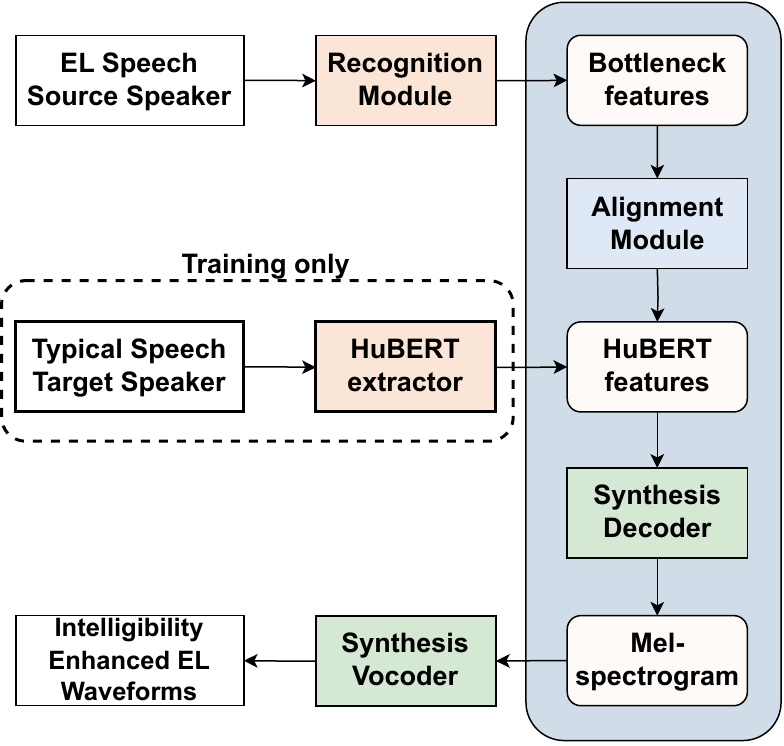} 
  \vspace{-1.0em}
  \caption{An overview of the proposed framework. The framework contains three main modules to convert from EL to typical speech: the recognition, alignment, and synthesis modules. Note that each module is trained separately.}
  \label{fig:archi}
\end{figure}

An overview of the entire framework can be seen in Fig. \ref{fig:archi}.  We detail the task of each module of the proposed framework below.

\subsection{Recognition module}
The recognition module uses a linguistic encoder, which uses the bottleneck features (BNFs) 
from an ASR encoder to extract the linguistic information. In our previous work \cite{el-intermediate-finetuning}, we showed that an effective approach to improving ASR model performance for EL speech is through a three-stage training framework. First, the model was pretrained on a large-scale typical speech dataset. Next, we fine-tuned the network on synthesized EL speech in an intermediate fine-tuning stage. Due to the limited data in training an EL speech synthesis model, the synthesized speech also contained lots of mispronunciations. However, we found that since the model used this stage to learn the EL speech characteristics instead of the linguistic information, it was sufficient enough for the synthetic EL speech to only represent the EL speech features. Finally, we fine-tuned the network on the ground truth EL speech to learn the linguistic features and decode at a high accuracy. We adopt this framework as the backbone of the recognition module.

The goal of the linguistic encoder now is to be robust enough to remove the speech type features from both typical and EL speech, while also accurately extracting linguistic information. With a unified representation, the performance of a pretraining and fine-tuning framework becomes robust to the speech type mismatches. Although this has been an easy task in typical VC, several ASR works have found developing speaker-independent models \cite{pretrain-deaf, pretrain-dysarthric-shor, pretrain-dysarthric-outperforming, pretrain-personalized} for atypical speakers a difficult task due to the high variance in their speech. A naive approach to resolve this would be to simply fine-tune the ASR model on both the EL and typical speech such that the model is not only optimized for EL speech. However, similar to previous works, we found that fine-tuning the ASR model on both types of speech at the same time causes performance degradations. 

To improve this, we simply introduce a speech type ID loss $L_{\text{SID}}$ during training. As our previous work discovered that the intermediate fine-tuning focuses on learning speech type identity features, we make the network learn both speech types during this stage. Let $X =\{X_{\text{TYP}}, X_{\text{EL}}\}$ be the training data, which is composed of a typical and an EL dataset $X_{\text{TYP}}$ and $X_{\text{EL}}$. The speech type ID loss $L_{\text{SID}}$ identifies whether the speaker is a typical or EL speaker and is optimized using a binary cross-entropy loss. Since we use both EL and typical data, we mask the outputs from the typical speech inputs during the calculation of the CTC/Attention losses $L_{\text{ctc}}$ and $L_{\text{attn}}$ \cite{ctc-attn}. The masking avoids making the model learn two highly variant types of speech, improving decoding performance. Through this approach, we effectively optimize the ASR model for EL speech, while also ensuring that it does not forget how to decode typical speech. We show in Eq. \ref{eq:asr} the detailed loss calculation during intermediate fine-tuning. After the intermediate fine-tuning stage, we fine-tune on $X_{\text{EL}}$ with the CTC/Attention losses $L_{\text{ctc}}$ and $L_{\text{attn}}$ \cite{ctc-attn} as usual.

\begin{equation}
    \label{eq:asr}
    L_{\text{ASR}} = L_{\text{SID}}(X) + L_{\text{ctc}}(X_{EL}) + L_{\text{attn}}(X_{EL})
\end{equation}

\subsection{Alignment module}
\label{alignment}
The alignment module resolves the intelligibility enhancement aspect. To improve intelligibility, the alignment module needs to fulfill two tasks. First, due to the temporal structure of EL speech, the model needs to increase the speaking rate similar to a typical speaker. Next, since EL speakers cannot produce certain phonemes, the alignment module also needs to correct the phoneme pronunciation. Similar to the baseline described in Section \ref{baseline}, we adopt the use of a Transformer \cite{transformer, vtn_journal} sequence-to-sequence model to resolve these issues. We also adopt the same fine-tuning procedure with synthetic data and then the target data due to its success. 

We improve this framework by using the BNFs produced by the recognition module as the inputs. These BNFs would further reduce the mismatches in speech type and speakers during pretraining and fine-tuning, as the linguistic encoder allows the alignment module to solely focus on modeling linguistic information. To further reduce the burden on the alignment network, we also use HuBERT as the output features. Aside from HuBERT providing dense linguistic information, using a variant of HuBERT with soft features \cite{hubert-soft} has also been found successful in removing speaker features and in cross-lingual settings, which would further improve the performance of the synthesis module described later.

Moreover, the TTS/AE pretraining described in Section \ref{baseline} was initially used in our baseline due to the unavailability of a large-scale parallel dataset; however, with the release of \cite{hificaptain}, we first verify whether parallel VC is indeed better. Although using parallel VC pretraining would directly model the fine-tuning task, this would not contain the speaker-independent properties of the TTS/AE pretraining, which might cause more degradations in the multiple fine-tuning stages due to the speech type and speaker mismatches. However, owing to the proposed framework focusing solely on linguistic features, we hypothetically remove this possibility.

\subsection{Synthesis module}
As our goal is to force the alignment module to focus only on modeling linguistic information, the task of synthesizing into the target speaker is placed on a synthesis module. Since the typical dataset used as a target speaker is also limited in size, we use a Diffusion model \cite{ho2020denoising} for this module, as this framework has been proven effective in synthesizing speech in a target speaker even in few-shot settings \cite{kim2022guided}. To improve the few-shot performance of the Diffusion model, similar to \cite{kim2022guided}, we first pretrain on a large-scale multi-speaker dataset with classifier-free guidance \cite{ho2022classifier} and use fixed speaker embeddings and HuBERT features as conditioning features. Then, we adapt the model to the few-shot data for another set of iterations. To train the model, we iteratively add noise for $N$ timesteps to the mel-spectrogram and predict the noise at timestep $n$ during training by using the noisy mel-spectrogram at $n-1$ as input along with the conditioning features. During inference, we pass in Gaussian noise and predict the mel-spectrogram after $N$ iterations. Finally, to synthesize the audio waveforms from the predicted mel-spectrograms, we use HiFiGAN (V1) \cite{hifigan} as the vocoder.

\vspace{-0.4em}
\section{Experimental Settings}

\subsection{Datasets}

\begin{table}[t]
\centering
    \footnotesize
    \caption{Total duration (in mins) and number of utterances for each split used in our experiment.}
    \label{tab:datasets}
\begin{tabular}{ccccccc}
\toprule
& \multicolumn{3}{c}{\textbf{Minutes}} & \multicolumn{3}{c}{\textbf{No. of Utterances}}\\
\cmidrule(lr){2-4}
\cmidrule(lr){5-7}
\textbf{Dataset} & \textbf{Train} & \textbf{Dev} & \textbf{Test} & \textbf{Train} & \textbf{Dev} & \textbf{Test} \\
\midrule
\midrule
TYPICAL & 4.38 & 2.14 & 2.17 & 116 & 40 & 40  \\
EL & 5.77 & 2.96 & 2.99 & 116 & 40 & 40 \\
\bottomrule
\end{tabular}
\end{table}

As the EL dataset is spoken in Japanese; thus, unless otherwise stated, the following datasets are also in Japanese. For the recognition module, we followed the same training framework as in \cite{el-intermediate-finetuning} to train a linguistic encoder. We first pretrained on a large-scale typical speech dataset containing around 2k hours of speech data \cite{laborotv}. Next, we fine-tuned the network on a total of 27k utterances of synthetic EL data and typical speech. Finally, we fine-tuned the network on our privately acquired EL speech data. We evaluated the performance of the linguistic encoder using the aforementioned EL data, and its parallel counterpart spoken by a typical speaker. We used a 116/40/40 split for the train, dev, and test data. We also conducted ablation studies on the performance when using a larger dataset from 15 simulated EL speakers (spoken by using an external electrolarynx) and their typical speech.

For the alignment module, we first pretrained our network on HiFiCaptain \cite{hificaptain}, a large-scale parallel dataset of typical speakers, containing around 18k utterances in total. We used the female speaker as the source and the male speaker as the target. Then, we used the same setup as in \cite{el_ma} where we first fine-tuned on synthetic EL, synthesized from text from the JSUT \cite{jsut} corpus. We then fine-tuned the model on our target parallel EL and typical speech data in Table \ref{tab:datasets}. Note that compared to our previous work in \cite{el_ma}, this current split is different, as we composed the evaluation data with longer utterances to show the effectiveness of the proposed method. We used the same 116/40/40 split used in the recognition module, so the test data is unseen by both the recognition and alignment modules.

For the synthesis decoder, we used the JVS dataset \cite{jvs}, a dataset containing 30 hours of speech from 100 speakers, to pretrain the model before fine-tuning it on the target typical speech. For the synthesis vocoder, we used a pretrained model on VCTK \cite{vctk}, an English dataset with 44 speakers of around 40 hours in total. 

\begin{table}[t]
\centering
    \footnotesize
    \caption{Resulting CER on both EL and typical speech with different training data setups.}
    \label{tab:recognition_module}
\begin{tabular}{lcc}
\toprule
\textbf{Training data and method} & \textbf{EL} & \textbf{TYPICAL} \\
\midrule
Speaker-dependent on typical speech  & 77.1 & 4.3  \\
Speaker-dependent on target EL speech \cite{el-intermediate-finetuning} & 15.9 & 61.5 \\
Fine-tuned on multiple EL and typical & 28.7 & 3.8 \\
Fine-tuned on EL and typical speech & 18.1 & 16.6 \\
\multirow{2}{*}{\makecell[l]{\begin{tabular}[l]{@{}l@{}}Proposed fine-tuned on EL and \\ typical with speech type ID loss only\end{tabular}}} & \multirow{2}{*}{\makecell[l]{\begin{tabular}[l]{@{}l@{}}16.2\end{tabular}}}  & \multirow{2}{*}{\makecell[l]{\begin{tabular}[l]{@{}l@{}}6.8\end{tabular}}} \\
\\
\midrule
\multirow{3}{*}{\makecell[l]{\textbf{\begin{tabular}[l]{@{}l@{}}Proposed fine-tuned on EL and \\ typical with speech type ID loss \\ and masking\end{tabular}}}} & \multirow{3}{*}{\makecell[l]{\textbf{\begin{tabular}[l]{@{}l@{}}13.8\end{tabular}}}} & \multirow{3}{*}{\makecell[l]{\textbf{\begin{tabular}[l]{@{}l@{}}6.0\end{tabular}}}} \\
\\
\\
\bottomrule
\end{tabular}
\end{table}

\begin{table*}[th!]
\centering
    \footnotesize
    \caption{Objective and subjective evaluation results on the synthesized speech from different systems, along with the ground truth recorded speech. MOS is calculated with a 95\% confidence interval. We detail the input and output features, along with the pretraining method used in the alignment module.}
    \label{tab:alignment_module}
\begin{tabular}{llllccccc}
\toprule
\textbf{(System) Description} & \textbf{Inputs} & \textbf{Outputs} & \textbf{Pretraining method} & \textbf{MCD ($\downarrow$)} & \textbf{CER ($\downarrow$)} & \textbf{F0 RMSE ($\downarrow$)} & \textbf{F0 CORR ($\uparrow$)} & \textbf{MOS ($\uparrow$)}   \\
\midrule
(1) Baseline \cite{el_ma} & mel & mel & TTS/AE & 7.78 & 35.0 & 51.19 & 0.30 & 2.42 $\pm$ 0.17  \\
(2) Baseline (ablation) & mel & mel & Parallel VC & 7.70 & 33.5 & \textbf{49.95} & 0.35 & 2.38 $\pm$ 0.18 \\
(3) Proposed & BNF & HuBERT & TTS/AE & 7.45 & 32.2 & 50.39 & 0.28 & 2.90 $\pm$ 0.17 \\
\textbf{(4) Proposed} & \textbf{BNF} & \textbf{HuBERT} & \textbf{Parallel VC} & \textbf{7.14} & \textbf{19.0}  & 52.41  & 0.29 & \textbf{3.25 $\pm$ 0.15} \\
(5) Proposed (ablation) & BNF & mel & Parallel VC & 7.54 & 29.1 & 51.16 & \textbf{0.37} & 2.78 $\pm$ 0.18 \\
\midrule
Ground truth & - & - & - & - & 4.3 & - & - & 4.85 $\pm$ 0.07 \\
\bottomrule
\end{tabular}
\end{table*}

\vspace{-1.5em}
\subsection{Model architecture}
The recognition model used a Conformer architecture \cite{conformer}, which has 12 layers for both the encoder and decoder, the same as in \cite{el-intermediate-finetuning}. On the other hand, the Transformer model in the alignment module has six layers for both the encoder and decoder, the same as in \cite{el_ma}. The diffusion model of the synthesis decoder was adopted from \cite{diffsinger} and uses 512-dimension channels to predict the noise between each timestep. To handle speech inputs, we replaced the text encoder with the BNF encoder as the conditioning features. To integrate speaker information, we used a pretrained WavLM model\footnote{\url{https://huggingface.co/microsoft/wavlm-base-plus-sv}} (which was fine-tuned for speaker verification) as speaker embeddings and fused it to each residual block using conditional layer normalization \cite{adaspeech2}. We set the number of diffusion steps $N$ to 100. No changes were made in HiFiGAN (V1).

\vspace{-1.0em}
\subsection{Evaluation metrics}
For objective evaluations, we measured the synthesis quality through metrics such as character error rate (CER), mel-cepstral distortion (MCD), log F0 root mean square error (F0 RMSE), and log F0 correlation (F0 CORR). For CER, we used the same Conformer model in Table \ref{tab:recognition_module} trained on the large-scale typical speech. For subjective evaluations, we recruited 15 native Japanese speakers to measure the naturalness of the synthesized speech using a 5-scale mean opinion score (MOS) test\footnote{Demo: \url{lesterphillip.github.io/icassp2024_el_sie}}.

\vspace{-1.0em}
\section{Results and Discussion}
\subsection{Validating the recognition module}
We first present how to develop a robust linguistic encoder. We investigate different training setups as shown in Table \ref{tab:recognition_module}. First, we see the difficulty in using a speaker-independent model, as optimizing on either EL or typical speech results in degradations in the other. Thus, using a model optimized on just EL speech degrades the large-scale pretraining stage on typical speech. Next, we see that fine-tuning with multiple EL and typical speakers to make the model more generalized, also does not have any effectiveness, caused by the high variance between these speakers. Finally, we show that simply fine-tuning the model on both EL and typical speech can be effective but not fully optimized, as there is still a performance gap from the speaker-dependent setups.

We show that our proposed method of using a speech type ID loss and masking the typical speech during CTC-Attention loss calculation makes the model learn to decode both EL and typical speech. This is because the model learns how to decode EL speech, while also not forgetting typical speech features learned during pretraining through the speech type ID loss. To verify this, removing the masking of typical speech results in slightly worse scores. Through this, we can decode both EL and typical speech at an accuracy similar to the speaker-dependent setups.

\subsection{Comparison of input/output features}
We show the effectiveness of the proposed linguistic encoder in this task. As seen in Table \ref{tab:alignment_module} our proposed method of using the BNF/HuBERT features (Sys. 4) can significantly improve the synthesized speech with a 16\% improvement in CER and 0.83 in naturalness score over Sys. 1, the baseline that uses mel-spectrograms as inputs. This proves our initial hypothesis that the proposed linguistic encoder can effectively remove speech type information while also extracting accurate linguistic information. We also conducted a study by using mel-spectrogram outputs. As shown in Sys. 5, using HuBERT instead of mel-spectrograms as outputs helps further stabilize the model, as it also contains dense linguistic information similar to the BNFs. Aside from this, Sys. 3 and 4 that used HuBERT features and the synthesis decoder had the top naturalness scores, further showing the effectiveness and necessity of a synthesis decoder over directly predicting the mel-spectrogram.

\subsection{Comparison of pretraining techniques}
In Section \ref{alignment}, we discussed that the TTS/AE pretraining also helps in resolving the speech type and speaker mismatches during pretraining and fine-tuning through its speaker-independent pretraining style \cite{vtn_journal}. However, upon comparing the baseline techniques, we observe Sys. 2 to have slightly better scores than Sys. 1 except in MOS. Thus, we prove that TTS/AE is not sufficient to create a speaker-independent property. Through the proposed approach in Sys. 4, we can directly model the fine-tuning task by using parallel VC pretraining, while also being able to implement a speaker-independent property by using BNF/HuBERT as input and output features, which reduces the mismatches during each fine-tuning stage. It is important to note that although Sys. 3 used both speaker-independent training styles, since the input (BNF) and output (HuBERT) features were different, the proposed method was not able to fully utilize the effectiveness of AE pretraining. Finally, we find that compared to the other systems, Sys. 4 has the highest F0 RMSE score and the second lowest F0 CORR score, showing that the proposed method truly allowed the alignment module to focus on modeling linguistic information, but caused a small tradeoff in modeling pitch. 

\vspace{-1.0em}
\section{Conclusions}
We proposed the use of robust linguistic encoders to remove speech features from both EL and typical speech. The major benefit that this brings is that it creates a unified representation for both EL and typical speech, reducing the speech type mismatches between each dataset in a pretraining and fine-tuning framework. The proposed method allows the model to focus on modeling intelligibility, where it outperforms the baseline with a 16\% improvement in CER and a 0.83 higher naturalness score. 

\noindent\textbf{Acknowledgements} This work was partly supported by AMED under Grant Number JP21dk0310114, Japan, and by JST CREST under Grant Number JPMJCR19A3.
\section{References}
\printbibliography

@ARTICLE{vc-intro,
  author={Sisman, Berrak and Yamagishi, Junichi and King, Simon and Li, Haizhou},
  journal={IEEE/ACM TASLP}, 
  title={An Overview of Voice Conversion and Its Challenges: From Statistical Modeling to Deep Learning}, 
  year={2021},
  volume={29},
  pages={132-157},
}

@article{electrolarynx,
author = {Mark I. Singer and Eric D. Blom},
title ={An Endoscopic Technique for Restoration of Voice after Laryngectomy},
journal = {Annals of Otology, Rhinology \& Laryngology},
volume = {89},
number = {6},
pages = {529-533},
year = {1980},
doi = {10.1177/000348948008900608},
    note ={PMID: 7458140},

URL = { 
        https://doi.org/10.1177/000348948008900608
    }
}

@inproceedings{project-euphonia,
  author={Robert L. MacDonald and Pan-Pan Jiang and Julie Cattiau and Rus Heywood and Richard Cave and Katie Seaver and Marilyn A. Ladewig and Jimmy Tobin and Michael P. Brenner and Philip C. Nelson and Jordan R. Green and Katrin Tomanek},
  title={{Disordered Speech Data Collection: Lessons Learned at 1 Million Utterances from Project Euphonia}},
  year=2021,
  booktitle={Proc. Interspeech},
  pages={4833--4837},
  doi={10.21437/Interspeech.2021-697}
}

@inproceedings{parrotron,
  title={Parrotron: An End-to-End Speech-to-Speech Conversion Model and its Applications to Hearing-Impaired Speech and Speech Separation},
  author={Fadi Biadsy and Ron J. Weiss and Pedro J. Moreno and D. Kanvesky and Ye Jia},
  booktitle={Proc. Interspeech},
  year={2019}
}

@INPROCEEDINGS{extending-parrotron,  author={Doshi, Rohan and Chen, Youzheng and Jiang, Liyang and Zhang, Xia and Biadsy, Fadi and Ramabhadran, Bhuvana and Chu, Fang and Rosenberg, Andrew and Moreno, Pedro J.},  booktitle={Proc. ICASSP},   title={{Extending Parrotron: An End-to-End, Speech Conversion and Speech Recognition Model for Atypical Speech}},   year={2021},  volume={},  number={},  pages={6988-6992},  doi={10.1109/ICASSP39728.2021.9414644}}

@inproceedings{conformer-parrotron, title	= {Conformer Parrotron: a Faster and Stronger End-to-end Speech Conversion and Recognition Model for Atypical Speech}, author={Zhehuai Chen and Bhuvana Ramabhadran and Fadi Biadsy and Xia Zhang and Youzheng Chen and Liyang Jiang and Andrea Chu and Rohan Doshi and Pedro Jose Moreno Mengibar}, booktitle={Proc. Interspeech}, year= {2021} }

@inproceedings{el_ma,
  title={{Two-stage training method for Japanese electrolaryngeal speech enhancement based on sequence-to-sequence voice conversion}},
  author={Ma, Ding and Violeta, Lester Phillip and Kobayashi, Kazuhiro and Toda, Tomoki},
  booktitle={Proc. SLT},
  pages={949--954},
  year={2023},
  organization={IEEE}
}

@INPROCEEDINGS{vtn-el-first,
  author={Yen, Ming-Chi and Huang, Wen-Chin and Kobayashi, Kazuhiro and Peng, Yu-Huai and Tsai, Shu-Wei and Tsao, Yu and Toda, Tomoki and Jang, Jyh-Shing Roger and Wang, Hsin-Min},
  booktitle={Proc. ASRU}, 
  title={Mandarin Electrolaryngeal Speech Voice Conversion with Sequence-to-Sequence Modeling}, 
  year={2021},
  volume={},
  number={},
  pages={650-657},
  doi={10.1109/ASRU51503.2021.9687908}}

@inproceedings{robust_encoder_atypical,
  author={Disong Wang and Songxiang Liu and Lifa Sun and Xixin Wu and Xunying Liu and Helen Meng},
  title={{Learning Explicit Prosody Models and Deep Speaker Embeddings for Atypical Voice Conversion}},
  year=2021,
  booktitle={Proc. Interspeech 2021},
  pages={4813--4817},
  doi={10.21437/Interspeech.2021-285}
}

@INPROCEEDINGS{laborotv,  
author={Ando, Shintaro and Fujihara, Hiromasa},  
booktitle={Proc. ICASSP},
title={{Construction of a Large-Scale Japanese ASR Corpus on TV Recordings}},
year={2021},
volume={},
number={},
pages={6948-6952},
doi={10.1109/ICASSP39728.2021.9413425}
}

@misc{hificaptain,
  author       =   {Takuma Okamoto and Yoshinori Shiga and Hisashi Kawai},
  title         =   {{Hi-Fi-CAPTAIN: High-fidelity and high-capacity conversational speech synthesis corpus developed by NICT}},
  published  =   {https://ast-astrec.nict.go.jp/en/release/hi-fi-captain/},
  year          =   {2023},
}

@article{jvs,
  title={{JVS corpus: free Japanese multi-speaker voice corpus}},
  author={Takamichi, Shinnosuke and Mitsui, Kentaro and Saito, Yuki and Koriyama, Tomoki and Tanji, Naoko and Saruwatari, Hiroshi},
  journal={arXiv preprint arXiv:1908.06248},
  year={2019}
}

@article{vctk,
title={{CSTR} {VCTK} corpus: {E}nglish multi-speaker corpus for {CSTR} voice cloning toolkit},
author={C.~Veaux, J.~Yamagishi, and K.~MacDonald},
journal={University of
  Edinburgh. The Centre for Speech Technology Research (CSTR)},
url={\url{http://dx.doi.org/10.7488/ds/1994}},
year={2012}
}

@INPROCEEDINGS{el-intermediate-finetuning,
  author={Violeta, Lester Phillip and Ma, Ding and Huang, Wen-Chin and Toda, Tomoki},
  booktitle={Proc. ICASSP}, 
  title={Intermediate Fine-Tuning Using Imperfect Synthetic Speech for Improving Electrolaryngeal Speech Recognition}, 
  year={2023},
  volume={},
  number={},
  doi={10.1109/ICASSP49357.2023.10095931}}

@inproceedings{pretrain-dysarthric-shor,
  author={Joel Shor and Dotan Emanuel and Oran Lang and Omry Tuval and Michael Brenner and Julie Cattiau and Fernando Vieira and Maeve McNally and Taylor Charbonneau and Melissa Nollstadt and Avinatan Hassidim and Yossi Matias},
  title={{Personalizing ASR for Dysarthric and Accented Speech with Limited Data}},
  year=2019,
  booktitle={Proc. Interspeech},
  pages={784--788},
}

@inproceedings{pretrain-dysarthric-outperforming,
  author={Jordan R. Green and Robert L. MacDonald and Pan-Pan Jiang and Julie Cattiau and Rus Heywood and Richard Cave and Katie Seaver and Marilyn A. Ladewig and Jimmy Tobin and Michael P. Brenner and Philip C. Nelson and Katrin Tomanek},
  title={{Automatic Speech Recognition of Disordered Speech: Personalized Models Outperforming Human Listeners on Short Phrases}},
  year=2021,
  booktitle={Proc. Interspeech},
  pages={4778--4782},
}

@article{pretrain-deaf,
  title={{An Analysis of Personalized Speech Recognition System Development for the Deaf and Hard-of-Hearing}},
  author={Violeta, Lester Phillip and Toda, Tomoki},
  journal={Proc. APSIPA},
  year={2023}
}

@INPROCEEDINGS{pretrain-personalized,
  author={Tobin, Jimmy and Tomanek, Katrin},
  booktitle={Proc. ICASSP}, 
  title={{Personalized Automatic Speech Recognition Trained on Small Disordered Speech Datasets}}, 
  year={2022},
  volume={},
  number={},
  pages={6637-6641},
  doi={10.1109/ICASSP43922.2022.9747516}}

@article{hifigan,
  title={Hi{F}i-{GAN}: Generative adversarial networks for efficient and high fidelity speech synthesis},
  author={Kong, Jungil and Kim, Jaehyeon and Bae, Jaekyoung},
  journal={Proc. NeurIPS},
  pages={17022--17033},
  year={2020}
}

@inproceedings{conformer,
  author={Anmol Gulati and James Qin and Chung-Cheng Chiu and Niki Parmar and Yu Zhang and Jiahui Yu and Wei Han and Shibo Wang and Zhengdong Zhang and Yonghui Wu and Ruoming Pang},
  title={{Conformer: Convolution-augmented Transformer for Speech Recognition}},
  year=2020,
  booktitle={Proc. Interspeech},
  pages={5036--5040},
  doi={10.21437/Interspeech.2020-3015}
}

@article{transformer,
  title={Attention is all you need},
  author={Vaswani, Ashish and Shazeer, Noam and Parmar, Niki and Uszkoreit, Jakob and Jones, Llion and Gomez, Aidan N and Kaiser, {\L}ukasz and Polosukhin, Illia},
  journal={Proc. NeurIPS},
  volume={30},
  year={2017}
}

@misc{jsut,
  doi = {10.48550/ARXIV.1711.00354},
  url = {https://arxiv.org/abs/1711.00354},
  author = {Sonobe, Ryosuke and Takamichi, Shinnosuke and Saruwatari, Hiroshi},
  title = {{JSUT corpus: free large-scale Japanese speech corpus for end-to-end speech synthesis}},
  publisher = {arXiv},
  year = {2017},
  copyright = {Creative Commons Attribution Share Alike 4.0 International}
}

@inproceedings{ctc-attn,
  title={Joint CTC-attention based end-to-end speech recognition using multi-task learning},
  author={Kim, Suyoun and Hori, Takaaki and Watanabe, Shinji},
  booktitle={Proc. ICASSP},
  pages={4835--4839},
  year={2017},
}

@ARTICLE{vtn_journal,
  author={Huang, Wen-Chin and Hayashi, Tomoki and Wu, Yi-Chiao and Kameoka, Hirokazu and Toda, Tomoki},
  journal={IEEE/ACM TASLP}, 
  title={Pretraining Techniques for Sequence-to-Sequence Voice Conversion}, 
  year={2021},
  volume={29},
  pages={745-755},
}

@misc{adaspeech2,
      title={AdaSpeech 2: Adaptive Text to Speech with Untranscribed Data}, 
      author={Yuzi Yan and Xu Tan and Bohan Li and Tao Qin and Sheng Zhao and Yuan Shen and Tie-Yan Liu},
      year={2021},
      eprint={2104.09715},
      archivePrefix={arXiv},
      primaryClass={cs.SD}
}

@article{hubert,
  title={Hubert: Self-supervised speech representation learning by masked prediction of hidden units},
  author={Hsu, Wei-Ning and Bolte, Benjamin and Tsai, Yao-Hung Hubert and Lakhotia, Kushal and Salakhutdinov, Ruslan and Mohamed, Abdelrahman},
  journal={IEEE/ACM TASLP},
  volume={29},
  pages={3451--3460},
  year={2021},
  publisher={IEEE}
}

@inproceedings{hubert-soft,
    author={van Niekerk, Benjamin and Carbonneau, Marc-André and Zaïdi, Julian and Baas, Matthew and Seuté, Hugo and Kamper, Herman},
    booktitle={Proc. ICASSP}, 
    title={A Comparison of Discrete and Soft Speech Units for Improved Voice Conversion}, 
    year={2022}
}

@article{ho2020denoising,
  title={Denoising diffusion probabilistic models},
  author={Ho, Jonathan and Jain, Ajay and Abbeel, Pieter},
  journal={Proc. NeurIPS},
  volume={33},
  pages={6840--6851},
  year={2020}
}

@inproceedings{ho2022classifier,
  title={Classifier-free diffusion guidance},
  author={Ho, Jonathan and Salimans, Tim},
  booktitle={Proc. NeurIPS},
  year={2021},
}

@article{kim2022guided,
  title={{Guided-TTS 2}: A diffusion model for high-quality adaptive text-to-speech with untranscribed data},
  author={Kim, Sungwon and Kim, Heeseung and Yoon, Sungroh},
  journal={arXiv preprint arXiv:2205.15370},
  year={2022}
}

@article{diffsinger,
  title={Diffsinger: Singing voice synthesis via shallow diffusion mechanism},
  author={Liu, Jinglin and Li, Chengxi and Ren, Yi and Chen, Feiyang and Liu, Peng and Zhao, Zhou},
  journal={arXiv preprint arXiv:2105.02446},
  volume={2},
  year={2021}}

@ARTICLE{liu2021bnfvc,
  author={Liu, Songxiang and Cao, Yuewen and Wang, Disong and Wu, Xixin and Liu, Xunying and Meng, Helen},
  journal={IEEE/ACM TASLP}, 
  title={Any-to-Many Voice Conversion With Location-Relative Sequence-to-Sequence Modeling}, 
  year={2021},
  volume={29},
  number={},
  pages={1717-1728},
  doi={10.1109/TASLP.2021.3076867}
}

\end{document}